# Deep-Tissue Anatomical Imaging of Mice Using Carbon Nanotube Fluorophores in the Second Near Infrared Window

Kevin Welsher*, Sarah P. Sherlock* and Hongjie Dai

Department of Chemistry, Stanford University, Stanford, CA 94305, USA

* These authors contributed equally.

† Correspondence and request for materials should be addressed to H.D., hdai@stanford.edu

**Abstract**

Fluorescent imaging in the second near infrared window (NIR II, 1-1.4 μm) holds much promise due to minimal autofluorescence and tissue scattering. Here, using well functionalized biocompatible single-walled carbon nanotubes (SWNTs) as NIR II fluorescent imaging agents, we performed high frame rate video imaging of mice during intravenous injection of SWNTs and investigated the path of SWNTs through the mouse anatomy. We observed in real-time SWNT circulation through the lungs and kidneys several seconds post-injection, and spleen and liver at slightly later time points. Dynamic contrast enhanced imaging through principal component analysis (PCA) was performed and found to greatly increase the anatomical resolution of organs as a function of time post-injection. Importantly, PCA was able to discriminate organs such as the pancreas which could not be resolved from real-time raw images. Tissue phantom studies were performed to compare imaging in the NIR II region to the traditional NIR I biological transparency window (700- 900 nm). Examination of the feature sizes of a common NIR I dye (indocyanine green, ICG) showed a more rapid loss of feature contrast and integrity with increasing feature depth as compared to SWNTs in the NIR II region. The effects of increased scattering in the NIR I versus NIR II region were confirmed by Monte Carlo simulation. In vivo fluorescence imaging in the NIR II region combined with PCA analysis may represent a powerful approach to high resolution optical imaging through deep tissues, useful for a wide range of applications from biomedical research to disease diagnostics.

**Introduction**

Fluorescence is an important imaging modality in life sciences and medicine(1) due to its ability to resolve features down to the diffraction limit and beyond.(2, 3) The benefit of higher resolution unfortunately comes at the cost of limited tissue penetration depth.(4) Absorption by biological tissue and water is one factor that leads to attenuation of signal proportional to the depth of the feature of interest. To overcome this problem, much research has focused on developing and implementing fluorescent probes in the "biological transparency window" near 800 nm (here, NIR I).(5) This NIR I transparency window is defined by a local minimum in the absorption spectrum of biological tissue, bounded by hemoglobin and water absorption (Fig. 1c) at shorter and longer wavelengths respectively. Many commercially available probes lie within this region, including commonly used cyanine dyes such as indocyanine green (ICG) and Cy 5.5,(6) as well as semiconductor quantum dots.(7)

While the traditional NIR probes outperform probes that emit at shorter wavelengths in the visible region, the definition of biological transparency window by absorption only leads to an incomplete assessment of the behavior of photons in turbid media. This is because photons emitted from a source embedded in turbid media, such as tissue, can be attenuated by both absorption and scattering events. Depth penetration in tissue is defined as $\delta = (3\mu_a(\mu_a + \mu_s'))^{-1/2}$, where $\mu_a$ is the absorption extinction coefficient, $\mu_s'$ ( $\sim \lambda^{-w}$) is the reduced scattering coefficient and $\delta$ is the resulting penetration depth. The exponent, $w$, depends on the size and concentration of scatterers in the tissue and ranges from 0.22 to 1.68 for different tissue (Fig. 1d).(8) Due to the relatively short wavelength of electromagnetic radiation emitted from traditional fluorophores (450 nm – 850 nm), coupled with the inverse wavelength of dependence of both Mie and Rayleigh scattering, the penetration depth can be greatly affected by the scattering nature of the tissue in question (Fig. 1d). The effect of scattering on the effective penetration depth of tissue has been observed experimentally(8, 9) and modeled via simulation,(10) with the consensus result being that the penetration depth can be maximized at wavelengths between 1 and 1.4 microns, called the second near infrared window (NIR II).

Currently, there is a dearth of available fluorophores emitting in this beneficial region. Quantum dots (QDs)(11) such as PbSe,(12) PbS,(13) and CdHgTe(14) are promising candidates for imaging in the NIR II, however, *in vivo* animal imaging in the NIR II region with QDs has not been carried out thus far. As an alternative, single-walled carbon nanotubes (SWNTs) have shown promise as fluorescent imaging agents in the NIR II both *in vitro* and *in vivo*.(15-18) In this work, we apply bright, biocompatible SWNTs as NIR II fluorophores for video rate *in vivo* imaging. Video rate fluorescent imaging of intravenously injected SWNTs allows crisp resolution of anatomical features via the application of dynamic contrast imaging through principal component analysis (PCA), as first developed by Hillman et al. using NIR I based

fluorophores.(19) Both the raw video and time based dynamic contrast enhanced images are used to investigate the pathway of SWNTs through the mouse anatomy up to 130 seconds post-injection. Finally, we perform imaging on mock tissue phantoms to determine the effect of using a NIR II versus a NIR I emitting fluorophore. These experimental results are backed up by Monte Carlo simulations, showing the beneficial effects of reduced tissue scattering on deeply embedded fluorophores emitting at longer wavelengths in the NIR II region.

**Results and Discussions**

**Video Rate NIR II Fluorescence Imaging of Mice**

Video rate imaging (see Movie S1,S2 in Supplementary Information) revealed that the path through the body for water soluble SWNTs (length ~ 200 – 500 nm, see Fig. S1 for an atomic force microscopy image)(18) coated by PEGylated phospholipid (DSPE-mPEG),(20) as monitored by their inherent NIR fluorescence, is similar to what is expected for tail-vein injection.(21) Following injection, the oxygen-poor blood travels to the heart and lungs to be oxygenated before being distributed throughout the body. This is evidenced by the high contrast of the lung features in the mice in both back and side views [3.5 seconds post-injection (p.i.), Fig 2a,e]. Following this spike in intensity, the lung contrast fades indicating the spread of the nanotube rich oxygenated blood to the rest of the circulatory system. Concurrent to the loss of lung contrast is a peak in kidney contrast, as shown in the images at 5.2 seconds p.i. (Fig. 2b,f) and region of interest (ROI) time courses (Fig. 2i,j, see Fig. S2 for ROI locations and raw ROI time course measurements with animal breathing). The increased signal at these early time points is a result of the nanotube-rich blood passing through the highly vascularized kidneys. Following the peak in kidney contrast, the kidney levels decrease towards its mean signal (Fig. 2i,j). The one-pass circulation time of blood in a mouse was previously determined to be ~15 seconds.(22) This time frame aligns with the appearance of a steady-state signal seen in ROI measurements in Figure 2. During the first 15 seconds of imaging, the SWNT-rich blood is distributed throughout the body. After the first-pass of SWNT-rich blood, the liver, lungs, muscle and kidneys all show fairly constant signal indicating consistent blood flow in these organs.

It should be noted that the strong signal does not necessarily indicate SWNT accumulation and retention in organs, but could be due to blood perfusion and microcirculation within organs, or temporary SWNTs extravasation prior to being re-released into the circulating blood. It is unlikely that the signal is solely due to SWNT accumulation and retention as the time frame of the video rate imaging (~2 minutes) is much shorter than the circulation half life of DSPE-mPEG functionalized SWNTs (~5 hours).(20) The strong kidney signal observed during imaging up to ~2 minutes post-injection does not support SWNT accumulation and retention as previous work has shown once nanotubes are no longer circulating, the level of SWNT accumulation in the kidneys is only ~1 % of the injected dose per gram (~1%ID/g) of tissue.(20) As a result, the peak in kidney signal is likely to be mainly due to SWNTs circulating in the blood stream shortly after injection. The SWNT signal in the head region in Fig. 2b is likely due to SWNT circulation rather than accumulation in the brain as little evidence exists that DSPE-mPEG functionalized SWNTs cross the blood-brain barrier.(20)

The behavior of organs in the reticuloendothelial system (RES), mainly the liver and spleen, are of particular interest in this study. It has been shown that clearance of SWNTs functionalized by DSPE-mPEG is through the RES system, with the liver and spleen showing uptake of ~35% and ~25% ID/g respectively at 1 day p.i.(20) The spleen is easily identified in the raw images at ~69 seconds p.i. (Fig. 2d,h). The spleen ROI shows continuously increasing signal over the course of the video imaging with no peak shown immediately following injection, contrary to what was seen from the lung and kidney ROIs (Fig. 2i,,j). While the physical meaning of this has not been fully investigated, the continual increase in fluorescence intensity could be attributed to SWNT extravasation or increasing microcirculation throughout the spleen. Previous studies have shown that the splenic marginal zone fills prior to the splenic red pulp.(23) Initial signal in the spleen could be due to filling of the marginal zone, followed by an increase in signal as SWNT-rich blood accumulates in the adjacent red pulp region.

The liver can be resolved in the raw images starting at ~17 seconds p.i. (Fig. 2c,g). The liver ROI shows similar behavior to both the lungs and kidney and does not follow the same behavior as the spleen despite the high SWNT accumulation at longer times as previously observed by *ex vivo* biodistribution studies(Fig. 2i,j).(20) It is likely that the liver has not begun to accumulate a significant amount of SWNTs at such early time points, and the signal seen is merely a reflection of the circulation of SWNT-rich blood through the organ.

**Anatomical Resolution through Principal Component Analysis**

In an effort to gain further anatomical resolution, PCA was applied to the time series of images based on the intrinsic fluorescence of SWNTs in NIR II. PCA is a common statistical processing method for compressing high dimensional data into a lower dimensional form by choosing only the highest variance components of the data set.(24) PCA is performed on groups of images by considering each pixel to be an observation which varies over the variable time. More information about PCA is provided in the Supplementary Information.

Dynamic contrast imaging was applied to observe behavior of organs which are not easily seen or resolved from nearby features in the raw images. As shown by Hillman et al.,(19) dynamic contrast imaging via PCA can be a powerful tool to achieve anatomical resolution of major organs. During initial circulation of a fluorophore, organ signal has the highest variation over time. Furthermore, different organs vary differently in time due to differences in blood volume flow rates for various organs(25), allowing them to be distinguished when proper statistical treatment is applied.

PCA of a time series of images was used as a tool to group pixels that vary similarly in time. As pixel variation within a single organ will be similar, PCA is able to group pixels belonging to the same organ. By defining axes in N-dimensional space which correlate to the highest variance and are mutually orthogonal, we translated the correlated time behavior of the pixels into spatial (in this case anatomical) resolution. Upon performing PCA analysis on the first 130 seconds post-injection, we observed four major organs of interest clearly (Fig. 3) due to their high, but distinct, variance in time. In addition, we believe we were able to resolve an additional organ, the pancreas (Fig. 3), which was largely overlapping with and blocked by the spleen and kidney and was not clearly distinguishable from the video-rate imaging alone. This demonstrates the unique ability of PCA to extract individual organ information from seemingly non-specific signal during imaging.

Since PCA is sensitive to large fluctuations, it is important to confirm that features observed are not artifacts created by breathing motions of the animal. To confirm that the breathing motions did not have an effect on the data, PCA was also run on a dataset where breathing frames were manually removed (see Movie S3 in Supplementary Information). Removal of breathing frames was equivalent to removing approximately one out of every three seconds of the original video. The results of PCA run on the edited dataset, Fig. S4, showed similar anatomical resolution as shown in Figure 3. In particular, this indicates that the pancreas signal observed is not an effect of splenic movement during breathing.

**Time Dependent PCA**

While PCA is a terrific tool for converting time correlation into spatial resolution, it comes at the cost of sacrificing temporal information. To get anatomical information from a time series of images, the images are compressed into orthogonal components. The insight of spatial resolution of features comes at the cost of temporal resolution. In an attempt to achieve both spatial and temporal resolution, we applied PCA over an expanding window over the time-series. In this way, the features from the early time points (lung, kidney) are maintained and features that appear later (liver, spleen, pancreas) change in appearance over different windows. This time-dependent PCA analysis was done to observe the prominence and clarity of individual organs over time (Fig. 4,S5). In agreement with the ROI time course data (Fig. 2i,j), the lung and kidneys are the prominent features when PCA is done over the initial time periods, as shown in the negative and combined PCA time-course images taken over the first 30 seconds p.i. (Fig. 4b,c;S5). Over the first 50 seconds p.i.. (Fig. 4a,c;S5), the positive and combined PCA time course images show increasing liver and spleen clarity as a function of time. It should be noted that the appearance of an organ in the time-course PCA images is an indication of high variance of those pixels in time. The appearance of the organ does not necessarily indicate uptake, nor the direction of the variance in time, be it increasing in intensity or decreasing.

One of the powers of PCA applied in this way is the ability to resolve features that cannot be seen from the raw images. The negative PCA time-course images show the blue pancreas feature becoming increasingly apparent over time (Fig. 4b). The feature first appears at 80 seconds p.i. and grows in size and clarity until 130 seconds p.i.. This contrasts with the positive PCA time-course images, which contain information about the RES organs (liver, spleen) that show different temporal behavior than the pancreas (Fig. 4a). The liver and spleen show up clearly at 50 seconds p.i. and increase in clarity with time. This difference in temporal behavior is a key piece of evidence that the pancreas and spleen features are in fact distinct. This temporal difference is further elucidated by an ROI time course taken over the positive pixels of the second principle component (spleen) and the negative pixels of the fourth principle component (pancreas) (Fig. S6).

Further, the combined PCA images show that the pancreas and spleen features are spatially distinct. Careful inspection of the combined PCA time-course images at 50 and 110 seconds p.i.. (Fig. 4c,S5) shows the appearance of a blue spot in between the spleen and kidney features. The spatial difference, combined with the temporal evidence discussed above, indicates that these two features are in fact distinct. The identification of this distinct feature was done by examining the mouse anatomy using Living Image software (Fig. S7). The only organ between the spleen and kidney is in fact the pancreas. The fact that the pancreas can be resolved beneath the spleen and kidney shows the depth penetration of the NIR II emitting SWNT probe and the power of time-dependent PCA to resolve buried features.

**Depth Penetration: NIR I versus NIR II**

The dynamic contrast images shown here provide evidence that SWNTs can be a useful tool for fast and sensitive fluorescent imaging. The contention of Hillman et al(19) was that the use of a NIR I dye (ICG) was beneficial due to the deeper tissue penetration and lower autofluorescence in this region. We propose the NIR II emitting dyes may be even more useful for this type of imaging, due to further reduction of autofluorescence (which is mostly confined to the visible)(26-32) and deeper tissue penetration. We further propose that the far red emission of SWNTs minimizes scattering, which is the main culprit in the loss of image fidelity and integrity. Whereas tissue absorption can be compensated for with higher excitation power and/or a brighter probe, scattering of light leads to loss of image information and is difficult to compensate or correct by any means.

Mock tissue phantom depth penetration studies were performed to investigate the effect of an absorbing and scattering medium (Intralipid®) on emitters in the NIR I (ICG) and NIR II (SWNT) (Fig.5). With increased depth, there is a qualitative and quantitative loss of image clarity and broadening of feature width. This is a direct result of the increased scattering as function of wavelength, which for Intralipid® is governed(33) by $\sim \lambda^{-2.4}$. Experimental studies have shown that most tissue follows the scattering behavior to varying degrees, with skin showing very high wavelength dependence (Fig. 1d).(8, 9) Intralipid® is known to have similar scattering properties and was chosen for this reason.(9)

The measured intensity profiles versus depth for NIR I and NIR II fluorophores were quite similar (Fig. 5c). This indicates that the depth penetration, as defined as a combination of both absorption and scattering, and measured as the surviving intensity of light, is similar in both cases.(4) Fitting an exponential function to these curves gives a characteristic depth penetration of 1.04 ± 0.04 and 0.97 ± 0.03 mm for ICG and SWNT respectively. The close similarity is due to the fact that for the NIR II emitter, the increased absorbance in water relative to the NIR I emitter, is compensated by a decreased scattering probability. This scattering effect is evidenced by the fact that the spread in the image increases significantly as a function of tissue depth and is more pronounced in the NIR I region than in the NIR II region (Fig. 5a,b). Due to the inverse wavelength dependence of scattering ($\sim\lambda^{-2.4}$ for Intralipid®), the NIR I region is characterized by higher albedo (i.e. high percentage of extinction due to scattering), and the loss of signal is correlated with a loss of feature contrast. The loss of image information is a direct result of a higher frequency of scattering events encountered by photons as they travel through turbid media. The scattering changes both the position and direction with which light leaves the medium relative to its starting trajectory, leading to a larger spread in the image features and confusion of feature information. The NIR II is a region of lower albedo, where the extinction is mainly dominated by absorption of water. In this region, the frequency of scattering events is lower, leading to better maintenance of feature information. Control experiments performed in a non-scattering medium (water, Fig. 5b) shows no spread of feature width with increasing depth. The loss of image integrity at shorter wavelengths is also observed in Monte Carlo simulations on a point source emitter (Fig. S8,S9), using water and Intralipid® as the major factors for absorption and scattering respectively. Despite the simple nature of the simulation, representing a single point rather than a macroscopic object, the qualitative trend is the same, with the NIR I showing greater loss of feature integrity with increasing depth. It should be noted that the extinction by an absorption event can be compensated for, especially in situations where the signal to background ratio is high (i.e. low autofluorescence, as in the NIR II), while a scattering event leads to confusion of the initial feature information including broadened feature sizes.

SWNTs show promise as a NIR II emitter for imaging in the low albedo domain. The NIR II region is not only characterized by low scattering, but also by low autofluorescence.(26) This benefit is further enhanced by a large Stokes shift between emission and excitation resonances ( $\geq$ 300 nm, see Fig. S10 for Photoluminescence excitation/emission (PLE) spectrum), leading to further suppression of endogenous autofluorescence.(18, 34) One issue is that the relatively low quantum yield (QY) of SWNTs may limit its future applications. Recent studies using separation and surface modification have led to somewhat enhanced QY,(35, 36) but further work needs to be done to apply these nanotubes to biological imaging. The development of new, brighter emitters such as semiconductor quantum dots and organic fluorophores in the NIR II will make exploitation of this advantageous region even more fruitful.

**Conclusion**

We have shown that SWNTs are useful fluorophores in the low albedo NIR II region. These fluorophores are bright enough to be imaged deep inside mice at high frame rate without excessive excitation power. Dynamic information from the raw images and ROI time courses, along with time dependent PCA analysis indicated that the circulation pathway of tail-vein injected SWNTs is from lungs to kidneys, followed by circulation in organs belonging to the RES system, represented in this work by the liver and spleen. PCA was also able to resolve the pancreas, which is buried beneath the spleen and kidney and was not observable in the raw time course images.

We have also shown that the NIR II emission window shows inherent advantages over the more commonly used NIR I window. Due to the lower albedo in this range, image contrast and feature sizes are better preserved with increased tissue thickness while low endogenous autofluorescence allows for easy discrimination between signal and background. This low background also allows for increased excitation power to compensate for absorption of water, and further increase the advantage of the NIR II window.

Due to the low background and low scattering, high anatomical resolution is possible in the NIR II region using PCA based dynamic contrast imaging. This dynamic contrast imaging may allow for many future applications, including the identification of tumors due to the differential blood flow rate of leaky tumor vessels. Furthermore, PCA imaging with high spatial resolution may be able to diagnose diseases which affect the size and shape of organs without the need for radioactive tags or MRI.

**Methods**

**Preparation of biocompatible SWNT fluorophores with high relative quantum yield.** The preparation of brightly fluorescent exchanged-SWNTs with high biocompatibility can be found in detail in Ref. (18) by using a surfactant exchange method to minimize damage to SWNTs. A fluorescence emission spectrum of the resulting SWNTs with DSPE-mPEG coating at 808 nm excitation is shown in Figure 1b. A photoluminescence excitation/emission (PLE) spectrum of the DSPE-mPEG SWNTs can be found in Fig. S10.

**Video Rate NIR II Imaging.** Video rate imaging was performed on a homebuilt setup consisting of a 2D InGaAs array (Princeton Instruments). The geometry of the imaging setup is shown in Figure 1a. The excitation light was provided by a fiber coupled 808 nm diode laser (RMPC Lasers), chosen to overlap with the traditional biological transparency window. The light was collimated by a 4.5 mm focal length collimator (ThorLabs) and filtered to remove unwanted radiation in the emission range. The excitation spot was a circle with a diameter of ~ 6 cm. The excitation power at the imaging plane was ~ 5W, leading to power density of ~140 mW/cm$^2$. Emitted light was passed through an 1100 nm longpass filter (ThorLabs FEL1100) and focused onto the detector by a lens pair consisting of two NIR achromats (200 mm and 75 mm, Thorlabs). The 1100 nm longpass filter was chosen to select the majority of the wavelength emission, while rejecting autofluorescence that may occur near the excitation band. The camera was set to expose continuously and images were acquired with LabVIEW software at highest possible frame rate. The exposure time for all images shown is 50 ms. There was a 19 ms overhead in the readout leading to an average time of 69 ms between consecutive frames. 2000 consecutive frames were collected leading to a total imaging time of 2 minutes and 18 seconds. For imaging, 5 female athymic nude mice were used and results shown are representative. The ideal concentration for injection and video rate imaging was determined to be 200 μL of ~500 nM SWNT solution (OD ~ 4 at 808 nm).(37)

**Mouse Handling and Injection** Female athymic nude mice were obtained from Harlan Sprague Dawley (Indianapolis, IN). All experiments were in accordance with Institutional Animal Care and Use Committee (IACUC) protocols. For SWNT injection, a 30 gauge catheter was inserted into the lateral tail vein allowing for bolus injection during the first frames of imaging.

**Dynamic Contrast Enhanced Images.** Dynamic contrast enhanced images were obtained in a similar fashion to the seminal work by Hillman et al.(19) All 2000 video frames were loaded into an array using MATLAB software and the princomp function was used to perform principal component analysis (PCA). PCA was performed using 150 frames evenly spaced over the entire data set. For "positive" images, the positive pixels for second, third and fourth principal components were assigned red, green and blue respectively and overlaid. For "negative" images, the negative pixels for second, third and fourth principal components were assigned red, green and blue respectively and overlaid. For "combined" images, the absolute value of the principal component scores were used. The construction of PCA images is shown in Fig. S3. Time-based dynamic contrast images were obtained by performing PCA over different periods of time. To conserve computing power, each time period was broken into 150 frames. For example, the 10 second dynamic contrast images were taken by performing PCA on the first 150 frames of the video. For the 30 second dynamic contrast images, PCA was performed over the first 450 frames, selecting every third frame.

**Tissue Phantom Depth Penetration Study.** Intralipid® was chosen as a mock tissue due to its similar scattering properties.(9, 10, 33, 38) A stock of 20% Intralipid® was diluted to make a 1% solution. Glass capillary tubes were filled with either SWNT or ICG solutions to act as mock fluorescent features. The bottom portion of capillary tubes was wrapped in black tape, leaving only a small portion on the top available for imaging. The capillaries were placed in a cylindrical dish and covered with different volumes of 1% Intralipid®. The depth of the capillary was calculated from the known area of the dish. Taping of the capillary tube improved depth measurements, as the depth was calculated from the exposed, fairly level portion of the capillary tube. Excitation was provided by a 785 nm diode laser (Renishaw) coupled to a 900 micron core optical fiber. Emitted light was filtered through an 850 nm longpass filter (Thorlabs FEL850) and imaged on the 2D InGaAs array described above. To obtain depth penetration information, average intensity was taken from the same ROI at various depths and plotted as a relative intensity for each probe. Plots of relative intensity for each probe prevent differences in probe quantum yields and extinction coefficients from affecting the comparison. To determine feature width, linear cross sections were taken from the images and fit to Gaussians using Origin's built in curve fitting. Linear cross sections taken from images, shown in Fig. S11, confirm that signal spreading is not due to phantom geometry or wavelength-dependent light interaction with the surface of the dish.

**Acknowledgements**

This work is supported by NIH-NCI 5R01CA135109-02 and CCNE-TR.

# References

1. Lakowicz, J. R. (2006) *Principles of fluorescence spectroscopy* (Springer, New York).
2. Rust, M. J., Bates, M. & Zhuang, X. (2006) Sub-diffraction-limit imaging by stochastic optical reconstruction microscopy (STORM). *Nat Methods* 3: 793.
3. Betzig, E., et al. (2006) Imaging Intracellular Fluorescent Proteins at Nanometer Resolution. *Science* 313: 1642-1645.
4. Splinter, R. & Hooper, B. A. (2007) *An introduction to biomedical optics* (Taylor & Francis, New York).
5. Britton, C. (1998) Near-Infrared Images Using Continuous, Phase-Modulated, and Pulsed Light with Quantitation of Blood and Blood Oxygenation. *Ann N Y Acad Sci* 838: 29-45.
6. Escobedo, J. O., Rusin, O., Lim, S. & Strongin, R. M. NIR dyes for bioimaging applications. *Curr Opin Chem Biol* 14: 64-70.
7. Michalet, X., et al. (2005) Quantum Dots for Live Cells, in Vivo Imaging, and Diagnostics. *Science* 307: 538-544.
8. Bashkatov, A. N. & et al. (2005) Optical properties of human skin, subcutaneous and mucous tissues in the wavelength range from 400 to 2000 nm. *J Phys D: Appl Phys* 38: 2543.
9. Tamara, L. T. & Suresh, N. T. (2001) Optical properties of human skin in the near infrared wavelength range of 1000 to 2200 nm. *J Biomed Opt* 6: 167-176.
10. Lim, Y. T., et al. (2003) Selection of quantum dot wavelengths for biomedical assays and imaging. *Mol Imaging* 2: 50-64.
11. Peng, X. G. (2009) An Essay on Synthetic Chemistry of Colloidal Nanocrystals. *Nano Res* 2: 425-447.
12. Wehrenberg, B. L., Wang, C. & Guyot-Sionnest, P. (2002) Interband and Intraband Optical Studies of PbSe Colloidal Quantum Dots. *J Phys Chem B* 106: 10634.
13. Bakueva, L., et al. (2004) PbS Quantum Dots with Stable Efficient Luminescence in the Near-IR Spectral Range. *Adv Mater* 16: 926-929.
14. Harrison, M. T., et al. (2000) Wet chemical synthesis and spectroscopic study of CdHgTe nanocrystals with strong near-infrared luminescence. *Mater Sci Eng, B* 69-70: 355.
15. Cherukuri, P., Bachilo, S. M., Litovsky, S. H. & Weisman, R. B. (2004) Near-Infrared Fluorescence Microscopy of Single-Walled Carbon Nanotubes in Phagocytic Cells. *J Am Chem Soc* 126: 15638-15639.
16. Jin, H., Heller, D. A. & Strano, M. S. (2008) Single-Particle Tracking of Endocytosis and Exocytosis of Single-Walled Carbon Nanotubes in NIH-3T3 Cells. *Nano Letters* 8: 1577.
17. Welsher, K., Liu, Z., Daranciang, D. & Dai, H. (2008) Selective Probing and Imaging of Cells with Single Walled Carbon Nanotubes as Near-Infrared Fluorescent Molecules. *Nano Letters* 8: 586-590.
18. Welsher, K., et al. (2009) A route to brightly fluorescent carbon nanotubes for near-infrared imaging in mice. *Nat Nano* 4: 773.
19. Hillman, E. M. C. & Moore, A. (2007) All-optical anatomical co-registration for molecular imaging of small animals using dynamic contrast. *Nat Photonics* 1: 526.
20. Liu, Z., et al. (2008) Circulation and long-term fate of functionalized, biocompatible single-walled carbon nanotubes in mice probed by Raman spectroscopy. *Proc Natl Acad Sci USA* 105: 1410-1415.
21. Hummel, K. P., Richardson, F. L. & Fekete, E. (1966), ed. Green, E. L. (Dover Publications, Inc, New York).

22. Debbage, P. L., et al. (1998) Lectin Intravital Perfusion Studies in Tumor-bearing Mice: Micrometer-resolution, Wide-area Mapping of Microvascular Labeling, Distinguishing Efficiently and Inefficiently Perfused Microregions in the Tumor. *J Histochem Cytochem* 46: 627-639.
23. Schmidt, E. E., MacDonald, I. C. & Groom, A. C. (1985) Microcirculation in mouse spleen (nonsinusal) studied by means of corrosion casts. *J Morphol* 186: 17-29.
24. Lay, D. C. (2003) *Linear Algebra and Its Applications* (Addison Wesley, Boston).
25. Davies, B. & Morris, T. (1993) Physiological Parameters in Laboratory Animals and Humans. *Pharm Res* 10: 1093.
26. Aubin, J. E. (1979) Autofluorescence Of Viable Cultured Mammalian-Cells. *J Histochem Cytochem* 27: 36-43.
27. Demchenko, A. P. (1986) *UV Spectroscopy Of Proteins* (Springer-Verlag, Berlin).
28. Longworth, J. W. (1971) in *Excited States Of Proteins And Nucleic Acids*, eds. Steiner, R. F. & Weinryb, I. (Plenum Press, New York), pp. 319-484.
29. Permyakov, E. A. (1993) *Luminescent Spectroscopy of Proteins* (CRC Press, Boca Raton).
30. Gafni, A. & Brand, L. (1976) Fluorescence Decay Studies Of Reduced Nicotinamide Adenine-Dinucleotide In Solution And Bound To Liver Alcohol-Dehydrogenase. *Biochemistry* 15: 3165-3171.
31. Sun, M., Moore, T. A. & Song, P.-S. (1972) Molecular luminescence studies of flavines. I. Excited states of flavines. *J Am Chem Soc* 94: 1730.
32. Benjamin, E., Zvi, M. & Yeshayahu, N. (1985) Fluorescence Spectral Changes of the Hematoporphyrin Derivative Upon Binding to Lipid Vesicles, Staphylococcus Aureus and Escherichia Coli Cells. *Photochem Photobiol* 41: 429-435.
33. Vanstaveren, H. J., et al. (1991) Light-Scattering In Intralipid-10-Percent In The Wavelength Range Of 400-1100 Nm. *Appl Opt* 30: 4507-4514.
34. Bachilo, S. M., et al. (2002) Structure-assigned optical spectra of single-walled carbon nanotubes. *Science* 298: 2361-236.
35. Crochet, J., Clemens, M. & Hertel, T. (2007) Quantum yield heterogeneities of aqueous single-wall carbon nanotube suspensions. *J Am Chem Soc* 129: 8058-8059.
36. Ju, S.-Y., Kopcha, W. P. & Papadimitrakopoulos, F. (2009) Brightly Fluorescent Single-Walled Carbon Nanotubes via an Oxygen-Excluding Surfactant Organization. *Science* 323: 1319-1323.
37. Kam, N. W. S., O'Connell, M., Wisdom, J. A. & Dai, H. (2005) Carbon nanotubes as multifunctional biological transporters and near-infrared agents for selective cancer cell destruction. *Proc Natl Acad Sci USA* 102: 11600-11605.
38. Stephen, T. F., et al. (1992) Optical properties of intralipid: A phantom medium for light propagation studies. *Laser Surg Med* 12: 510-519.

**Figures**

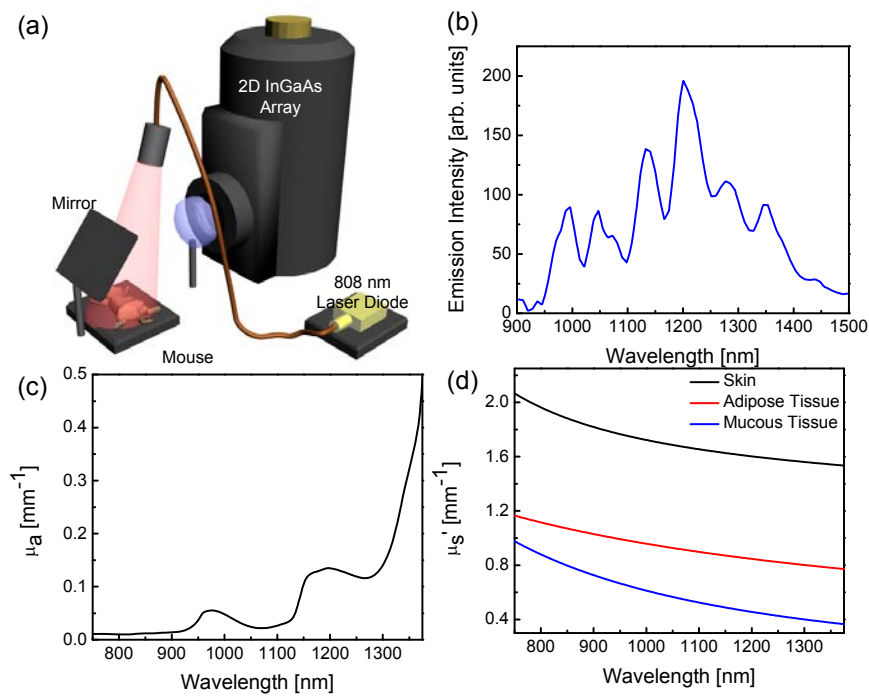

**Figure 1 | NIR II Imaging**. **(a)** Schematic of NIR II imaging setup. Anaesthetized mice are illuminated from above with 808 nm light. NIR fluorescence (1100 nm – 1700 nm) is filtered and imaged onto a 2D InGaAs array. **(b)** Fluorescence spectrum of biocompatible DSPE-mPEG functionalized SWNTs excited at 808 nm, showing several emission peaks spanning the NIR II region. **(c)** Absorption coefficient, $\mu_a$, of water, showing the increased absorption of water in the NIR II compared to the NIR I. **(d)** Reduced scattering coefficient, $\mu_s'$, for skin, adipose tissue and mucous tissue as derived in reference 8, all showing decreased scattering with increasing wavelength.

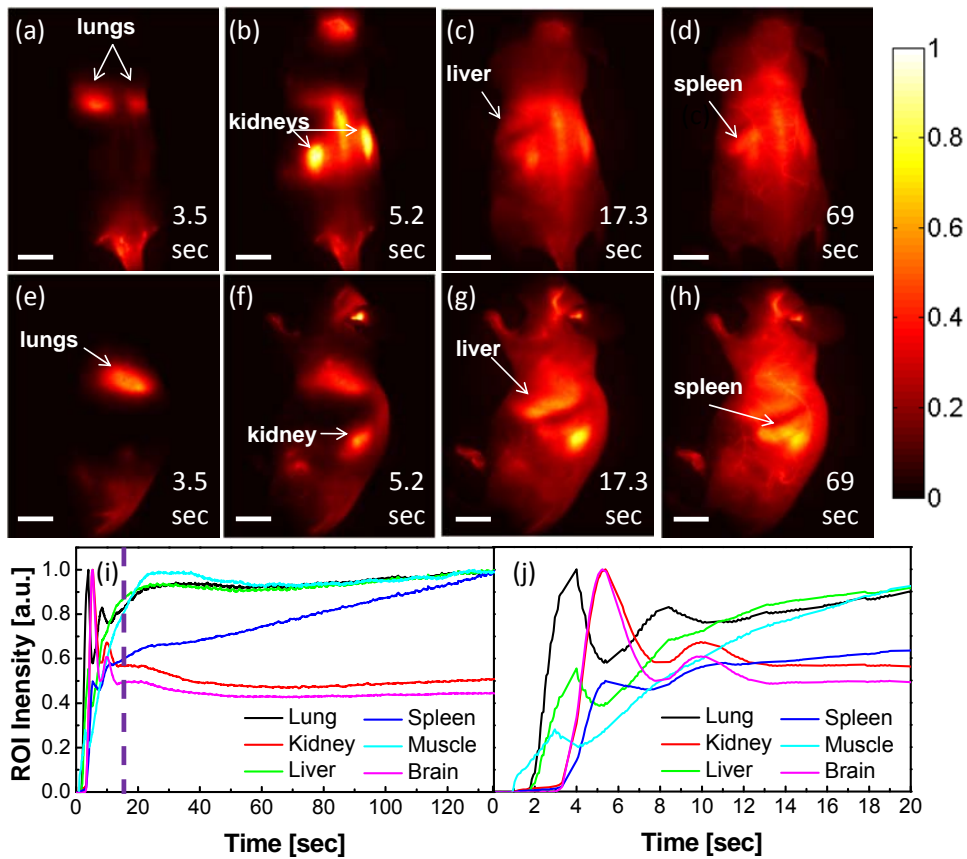

**Figure 2 | Video Rate Imaging of SWNTs in a Live Mouse.** Frames from video imaging of mice injected with SWNTs. At **(a,e)** 3.5 seconds following tail vein injection, the lungs are the dominant feature, corresponding to flow of the oxygen poor, SWNT rich blood to the lungs. **(b,f)** At 5.2 seconds, the SWNT rich blood flows through the highly vascularized kidney. **(c,g)** The liver becomes apparent at 17.3 seconds p.i.., while the **(d,h)** spleen becomes visible at 69 seconds p.i.. Scale bars in all images represent 1 cm. **(i,j)** Normalized region-of-interest (ROI) time courses over the organs in the raw images. For clarity, frames where the mouse is breathing were not included. The lungs, kidney and liver show large spikes shortly after injection (~5 sec) followed by a return to a steady state intensity within 20 seconds. The purple line in **(i)** shows the predicted time for blood to make one-pass through the body leading to a steady state SWNT signal in the body. The spleen shows a deviation from this behavior, showing no early spike and a monotonic increase in signal with increasing time.

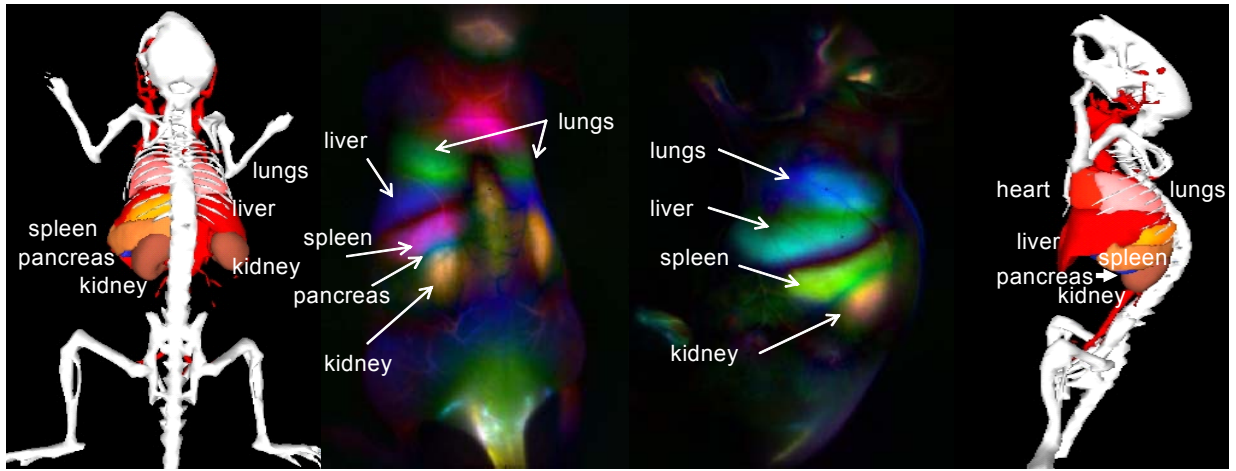

**Figure 3 | Dynamic Contrast Enhanced Imaging with SWNTs through PCA.** Principal component analysis images taken over the first 130 seconds following injection performed by taking every 150 evenly spaced frames out of the 2000 frame data set. Major features observed belong to the lungs, liver, kidney and spleen. Of note is the appearance of the pancreas in the interstitial space between the kidney and spleen (see text for details). This feature is not observable in the raw time course images.

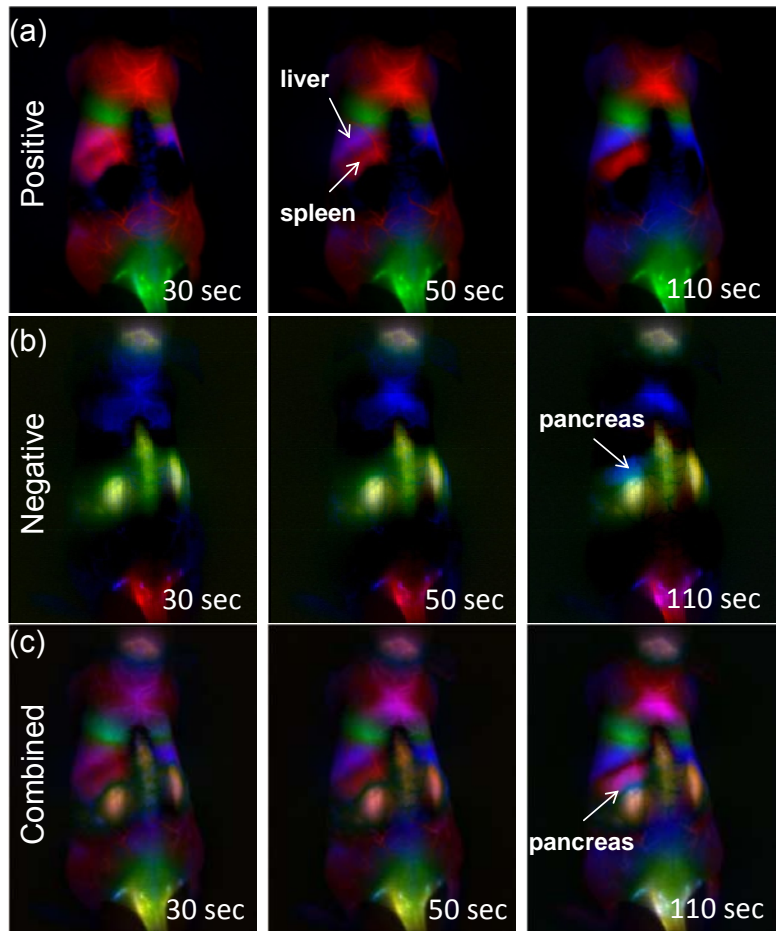

**Figure 4 | Time dependent PCA Images of Mouse Anatomy with SWNTs. (a)** Positive pixels of time dependent PCA, showing the liver and spleen of the RES. The liver and spleen show increasing clarity as a function of time. **(b)** Negative pixels from time dependent PCA showing kidney features at 30, 50 and 110 seconds. The pancreas appears as a blue spot above the left kidney in the 110 second image. **(c)** Absolute value of pixels from time dependent PCA analysis, showing increased clarity of liver and spleen as a function of time and a distinct blue pancreas feature at 110 seconds.

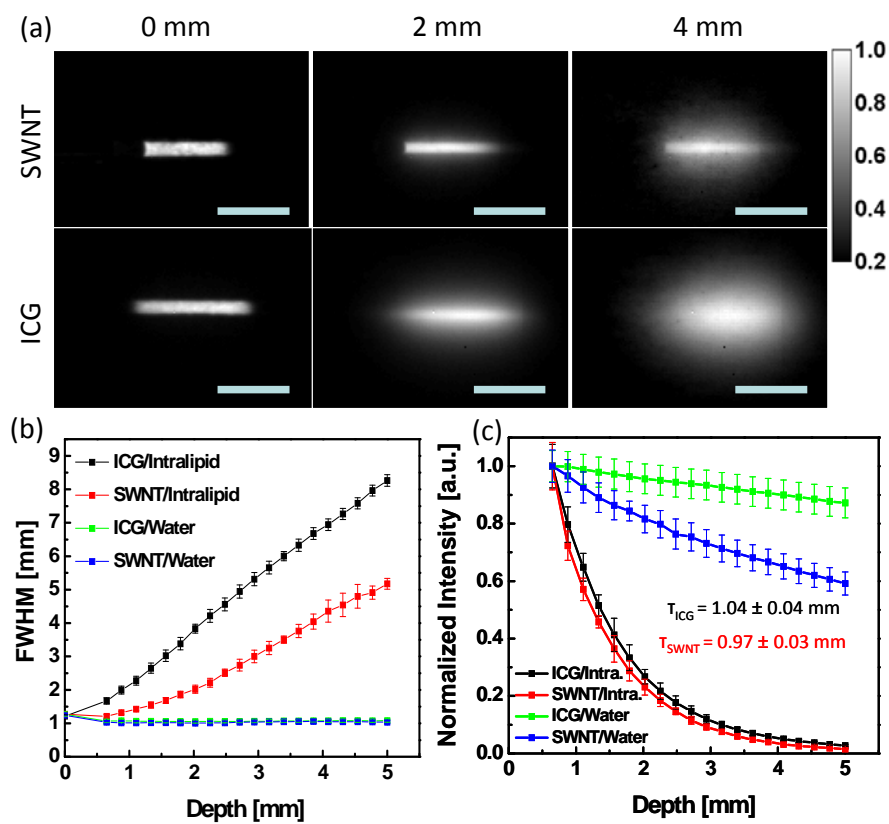

**Figure 5 | Tissue Phantom Study of the Depth Penetration of SWNTs and ICG. (a)** Fluorescence images of capillaries of SWNTs (NIR II) and ICG (NIR I) at depths of 0, 3 and 5 mm in Intralipid® excited at 785 nm. The SWNT sample shows less feature spread than that of the ICG sample. Scale bars represent 1.5 cm. **(b)** Feature width of SWNT and ICG capillary images as a function of depth in Intralipid®, showing increased loss of feature integrity for the NIR I emitting ICG compared to the NIR II emitting SWNT. Control experiments performed in water show no change in feature size for both ICG and SWNT. Error bars are derived from the uncertainty in the fitting of feature width. **(c)** Intensity decay of ICG and SWNT as a function of depth in Intralipid® and water. Despite the greater absorption of water in NIR II, the decay of signal in the Intralipid® phantom is similar for both ICG and SWNT, showing exponential decay depths of $1.04 \pm 0.04$ and $0.97 \pm 0.03$ mm respectively.

Supporting Information for:

Deep-Tissue Anatomical Imaging of Mice Using Carbon Nanotube Fluorophores in the Second Near Infrared Window

Kevin Welsher*, Sarah P. Sherlock* and Hongjie Dai

Department of Chemistry, Stanford University, Stanford, CA 94305, USA

**Principal Component Analysis (PCA)**

PCA is a common statistical processing method for compressing high dimensional data into a lower dimensional form by choosing only the highest variance components of the data set.(2) PCA is performed on groups of images by considering each pixel to be an observation which varies over the variable time. The data is N dimensional, where N is the number of time points in the series of images. The $p \times N$ matrix (where p is the number of pixels in the image) is converted to a covariance matrix by multiplying by its transpose and subtracting the time averaged mean values of each pixel. This covariance matrix is then diagonalized to find eigenvalues (variances) and eigenvectors (components). The eigenvectors point in the direction of greatest variance for a given component. The first component has the highest variance, similar to a weighted mean over all the images. The second component is in the direction of next highest variance which is orthogonal to the first component (Fig.S3). The third component (Fig.S3) is orthogonal to both the first and second components and so on. As a linear combination of the initial images, each principal component is made up of positive and negative pixels (Fig.S3), both of which indicate high variance but with different behavior along that particular component axis.

**Detailed Methods**

**Preparation of biocompatible SWNT fluorophores with high relative quantum yield.** The preparation of brightly fluorescent exchanged-SWNTs with high biocompatibility can be found in detail in Ref. (1) by using a surfactant exchange method to minimize damage to SWNTs. Briefly, raw HiPCO SWNTs (Unidym) were suspended in 1 wt % sodium cholate hydrate in water by 1 hour of bath sonication. This suspension was ultracentrifuged at 300,000g to remove bundles and other large aggregates. The supernatant was retained and 1 mg/mL of DSPE-mPEG (1,2-distearoyl-*sn*-glycero-3-phosphoethanolamine-N-[methoxy(polyethylene glycol)5000] (Laysan Bio) was added. The resulting suspension was dialyzed against a 3500 MCWO membrane (Fisher) with a minimum of six water changes and a minimum of two hours between water changes. As a final step, the suspension was ultracentrifuged again for 1 hour at 300,000 g to remove any aggregates. A fluorescence emission spectrum of the resulting SWNTs with DSPE-mPEG coating at 808 nm excitation is shown in Figure 1b. A photoluminescence excitation/emission (PLE) spectrum of the DSPE-mPEG SWNTs can be found in Fig. S10.

**Video Rate NIR II Imaging.** Video rate imaging was performed on a homebuilt setup consisting of a 2D InGaAs array (Princeton Instruments). The geometry of the imaging setup is shown in Figure 1a. The excitation light was provided by a fiber coupled 808 nm diode laser (RMPC Lasers). This wavelength was chosen to overlap with the traditional biological transparency window. It should be noted that other excitation/emission combinations are possible further into the NIR II, with larger diameter nanotubes exhibiting excitation and emission bands beyond 900 nm and 1500 nm respectively.(3) These wavelengths may have further reduced scattering, but an analysis of these regions is beyond the scope of this work. The light was collimated by a 4.5 mm focal length collimator (ThorLabs) and filtered to remove unwanted radiation in the emission range. The excitation spot was a circle with a diameter of ~ 6 cm. The excitation power at the imaging plane was ~ 5W, leading to power density of ~140 mW/cm$^2$. Emitted light was passed through an 1100 nm longpass filter (ThorLabs FEL1100) and focused onto the detector by a lens pair consisting of two NIR achromats (200 mm and 75 mm, Thorlabs). The 1100 nm longpass filter was chosen to select the majority of the wavelength emission, while rejecting autofluorescence that may occur near the excitation band. The camera was set to expose continuously and images were acquired with LabVIEW software at highest possible frame rate. The exposure time for all images shown is 50 ms. There was a 19 ms overhead in the readout leading to an average time of 69 ms between consecutive frames. 2000 consecutive frames were collected leading to a total imaging time of 2 minutes and 18 seconds. For imaging, 5 female athymic nude mice were used and results shown are representative. The ideal concentration for injection and video rate imaging was determined to be 200 μL of ~500 nM SWNT solution (OD ~ 4 at 808 nm).(4) It was observed that higher concentration lead to a loss of feature clarity, whereas lower concentration or injection volume led to lower than desired signal to noise.

**Mouse Handling and Injection** Female athymic nude mice were obtained from Harlan Sprague Dawley (Indianapolis, IN) and were housed at Stanford Research Animal Facility in accordance with Institutional Animal Care and Use Committee (IACUC) protocols. During imaging mice were anaesthetized by inhalation of 2% isoflurane with oxygen. For SWNT injection, a 30 gauge catheter was inserted into the lateral tail vein allowing for bolus injection during the first frames of imaging.

**Monte Carlo Simulation.** Simulations were performed following the procedure from Ref. (5) using MATLAB. The simulation considered photon packets of starting weight $W$ emitted from a point source embedded in a turbid medium. The emission angle was limited to $\pm 30°$ to conserve computing power. Simulation of a point source with limited emission angle could give quantitatively different results as obtained in our phantom experiments, but the qualitative trends seem to agree. The photon packet travelled a distance of $d = -ln(RAND)/\mu_s'$ before encountering a scattering event. Scattering was considered to be isotropic with a uniform angular distribution. After each scattering event, the weight of each packet was reduced by a factor of $exp(-\mu_a \cdot d)$. Upon reaching the tissue/air interface, the packet underwent refraction according to Snell's Law. The displacement and direction of the packet at the interface was then projected onto to the image space using ray matrices derived from the actual experimental setup. The imaged light packets were binned into 10 micron steps over a 2 mm by 2 mm area and the resulting images plotted. The values for $\mu_a$ for water ($0.098 \pm .002$ mm$^{-1}$ at 800 nm, $0.140 \pm 0.02$ mm$^{-1}$ at 1300 nm) were measured using a UV-Vis-NIR (Varian Cary 6000i). The value of $\mu_s'$ for Intralipid® was obtained from the literature following the relation $\mu_s' = 16 \cdot \lambda^{-2.4}$, where $\lambda$ is the wavelength in microns. (6) This resulted in $\mu_s'$ values of 2.73 and 0.852 at 800 nm and 1300 nm respectively.

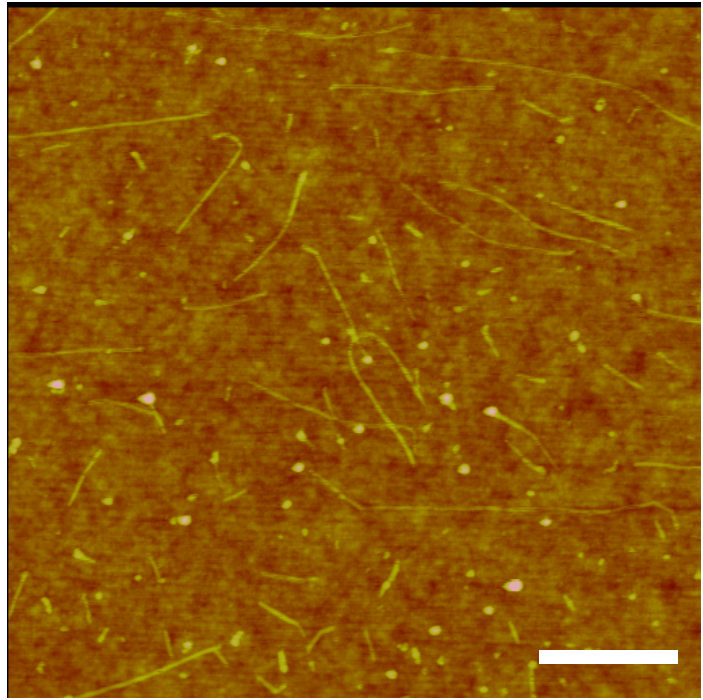

**Figure S1 | Atomic Force Microscopy.** AFM image of bright, biocompatible SWNTs in DSPE-mPEG deposited on a $SiO_2$ surface. Length ranges from 200 – 500 nm with a mean length of ~ 350 nm. Scale bar represents 200 nm.(1)

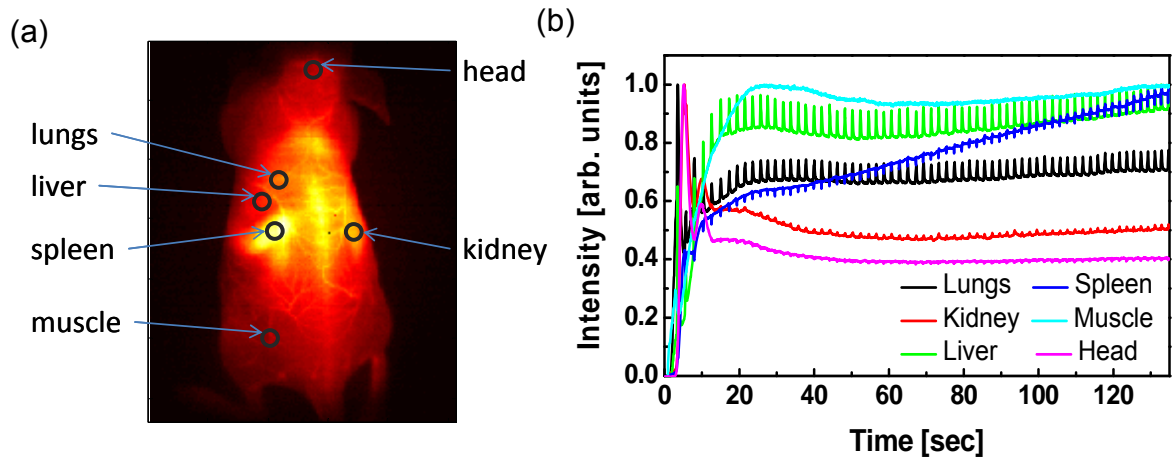

**Figure S2 | ROI Time Traces. (a)** Regions used to generate the normalized ROI time traces shown in Figure 2i,j. **(b)** ROI time traces obtained from the regions in (a) without removal of breathing frames.

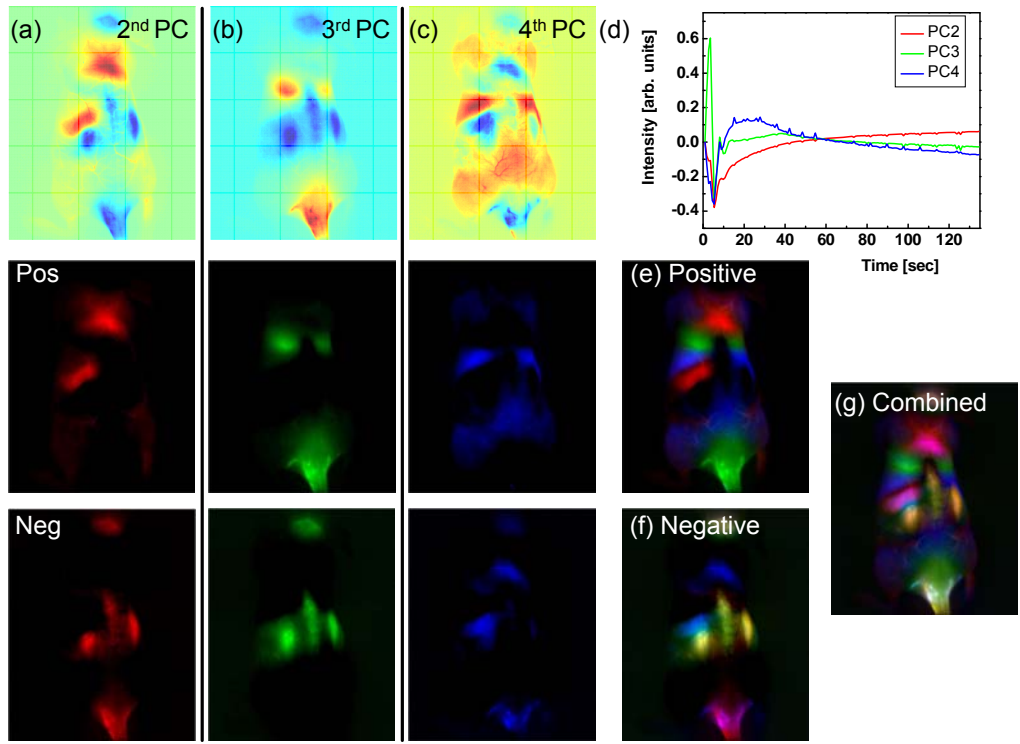

**Figure S3 | Building PCA Images.** Following PCA analysis by MATLAB, the **(a)** second (**red**), **(b)** third (**green**) and **(c)** fourth (**blue**) components are retained. **(d)** The principal component eigenvectors show how each principle component image is composed from the raw time course images. The y-axis reports the coefficient used for each x-axis time frame to make the linear combination of images for each principal component. **(e)** Positive image overlay. Only the positive pixels are retained by setting the negative pixels to zero. **(f)** For the negative images, the negative pixels are retained and inverted to make the composite image. **(g)** For the combined image, the absolute value of each pixel is plotted in an effort to retain all information.

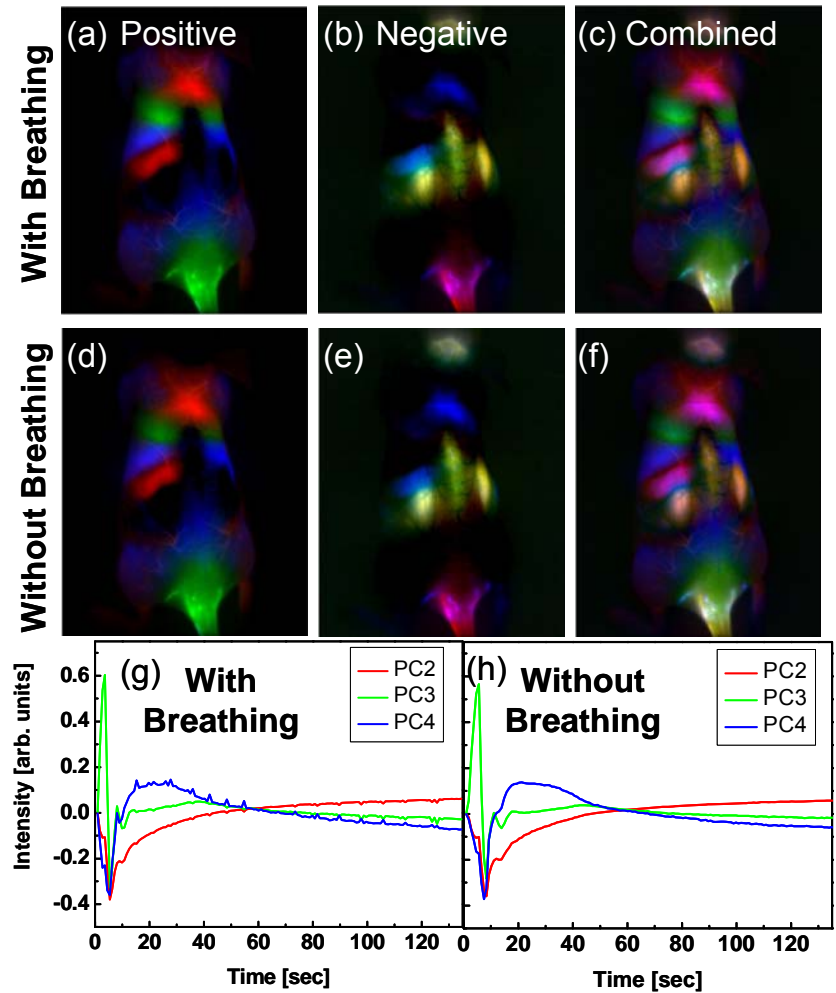

**Figure S4 | Effect of Breathing Movements on PCA.** **(a)** Positive pixels, **(b)** negative pixels and **(c)** a combined overlay obtained by performing PCA on a complete series ('With Breathing') of images as in Figure 3,4. To rule out breathing-related artifacts, frames involving movement of the animal were manually removed and PCA was performed on this edited series of images. The **(d)** positive pixels, **(e)** negative pixels and **(f)** combined overlay of the 'Without Breathing' series shows the same organs resolved in parts a-c. The similarity confirms that overlapping organs are not artifacts due to breathing motion of the animal. The principal component

eigenvectors of the **(g)** 'With Breathing' series shows a similar trend to the **(f)** 'Without Breathing' series. PCA for each dataset was performed on ~150 frames spanning the first ~130 seconds post-injection.

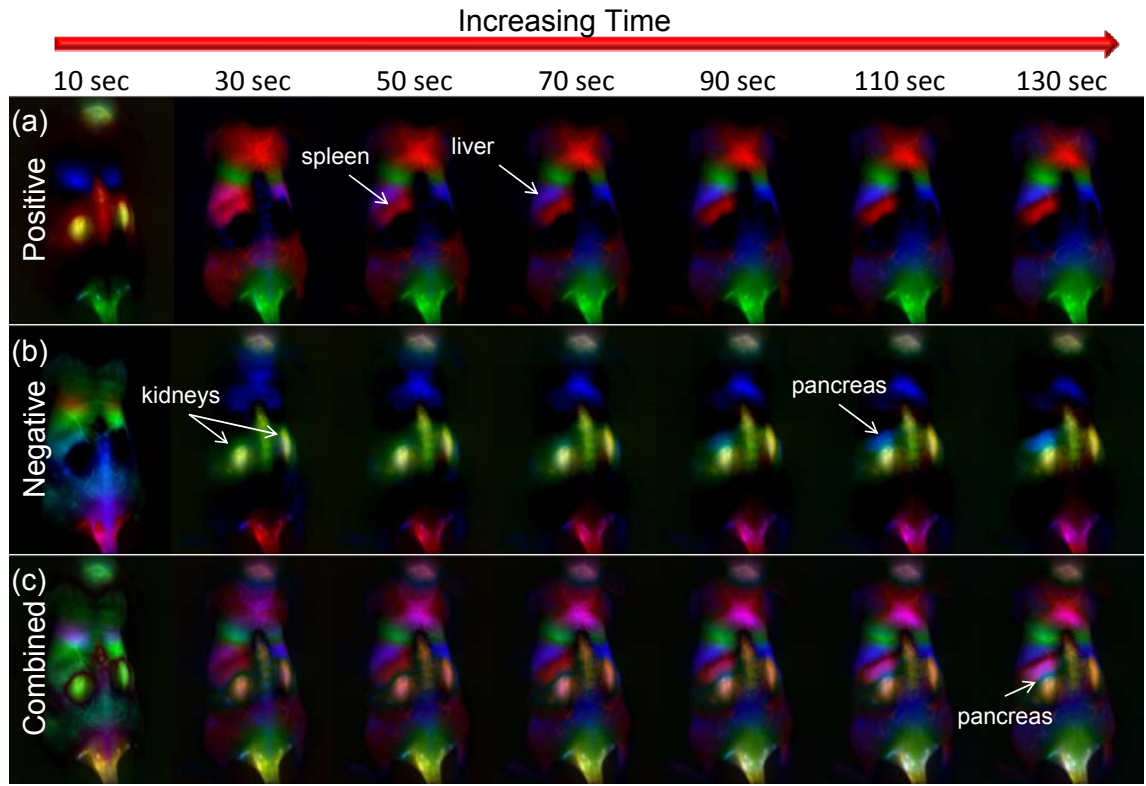

**Figure S5 | Time Dependent PCA Images.** **(a)** Positive only pixels from time dependent PCA. The liver and spleen show up at about 50 seconds and become clearer with increasing time. **(b)** Negative only pixels from time dependent PCA. The kidneys appear at 30 seconds and remain constant throughout the timecourse. The pancreas appears a blue feature above the left kidney at 90 seconds and increases in intensity with increasing time. **(c)** Absolute value of pixels from time dependent PCA, showing increased clarity of the RES organs (liver, spleen) with increasing time, as well as the appearance of a spatially distinct pancreas feature at 90 seconds.

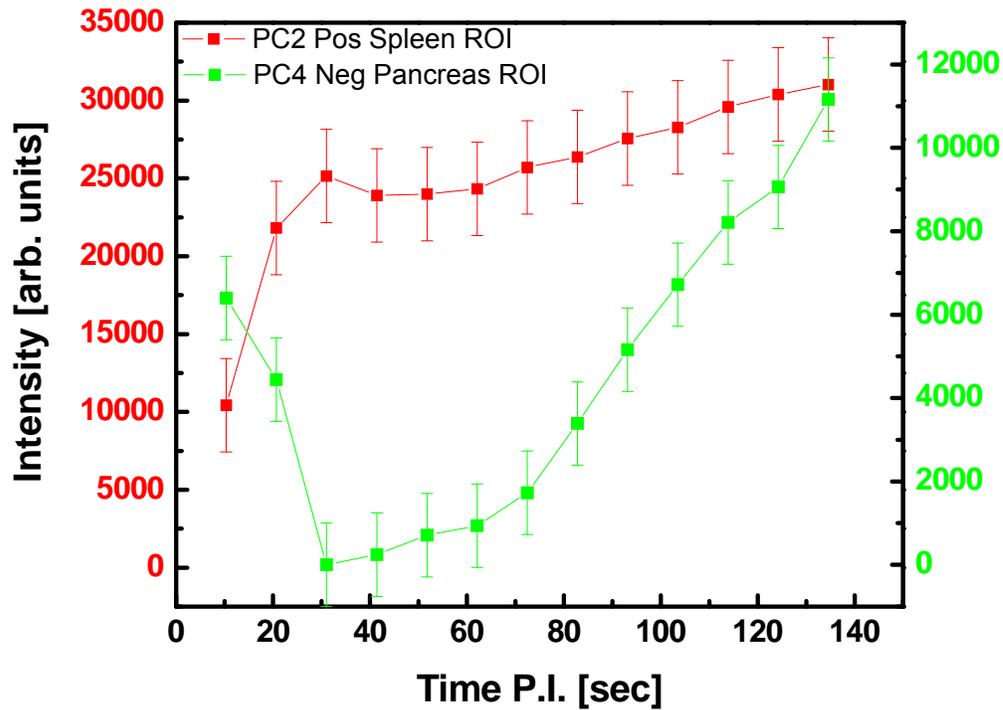

**Figure S6 | Time Dependent PCA ROI.** ROI time course of spleen and pancreas features from time dependent PCA, showing different behavior in time. For the red curve, ROIs were drawn over the spleen component in the positive pixels of the second principal component. An average intensity was taken at each time point in the time-course PCA data. The error bars represent the standard deviation of the intensity values in the ROI. The green curve was determined in a similar way, using the pancreas region of the negative pixels of the fourth principal component. The pancreas is not visible before 90 seconds, while the spleen is visible at 30 seconds. This temporal difference indicates that these two features belong to different organs.

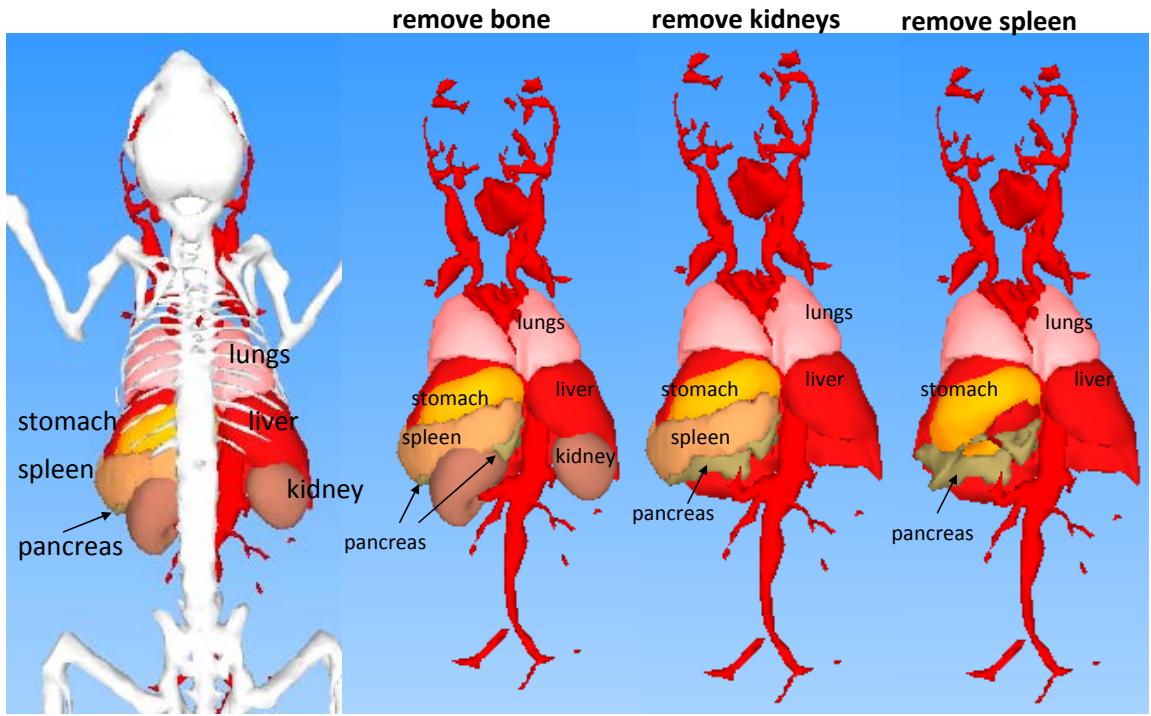

**Figure S7 | Pancreas Location.** Mouse anatomy showing the location of the pancreas, when the bone, kidney and spleen are successively removed.

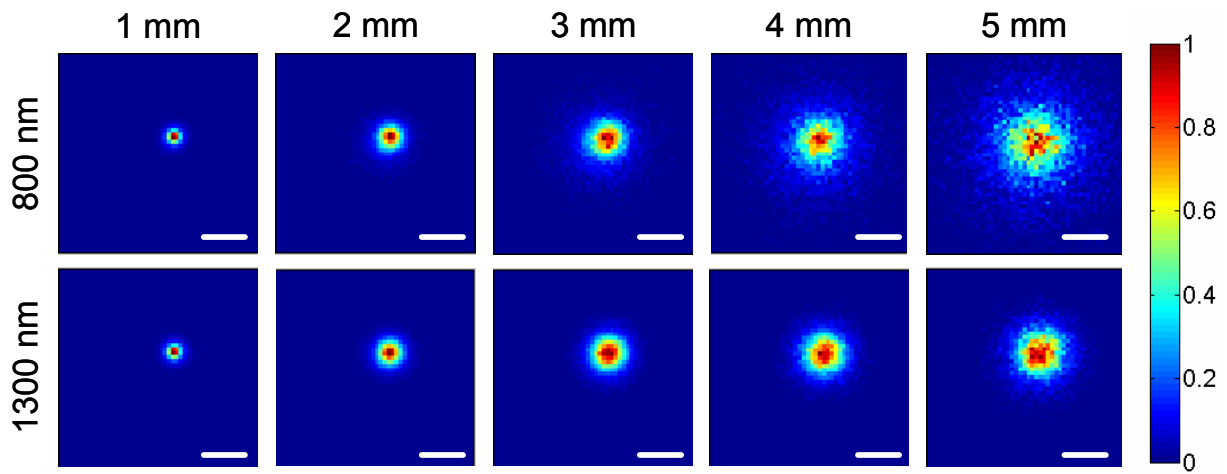

**Figure S8 | Monte Carlo Simulation.** Monte Carlo generated images (see Methods section in main text for details) as a function of emission wavelength and depth for a point source in Intralipid®. The computed image shows a greater spread in the feature size with depth for the 800 nm emitter (top row) compared to the 1300 nm emitter (bottom row), showing the benefit of imaging in the low albedo NIR II region. Scale bars represent 3 mm.

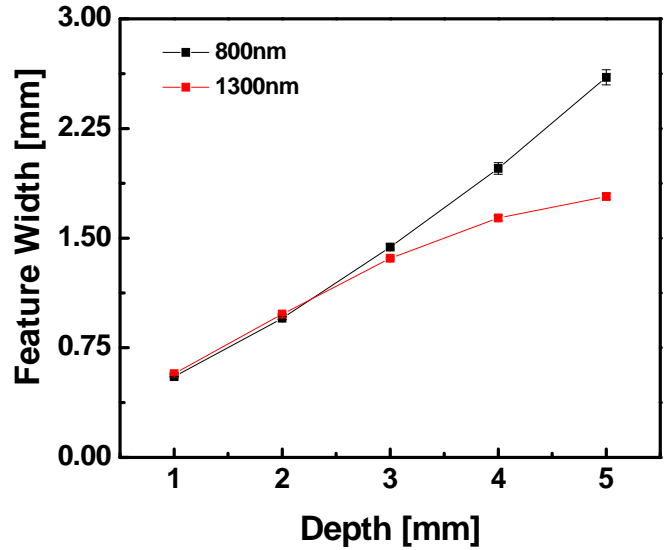

**Figure S9 | Monte Carlo Simulation Feature Spreading.** Feature width as a function of depth for the NIR I and NIR II, generated from the Monte Carlo simulation of emission from a point source embedded in Intralipid®. The simulation shows a greater loss of feature integrity for the NIR I. Error bars are derived from the uncertainty in the fitting of feature width.

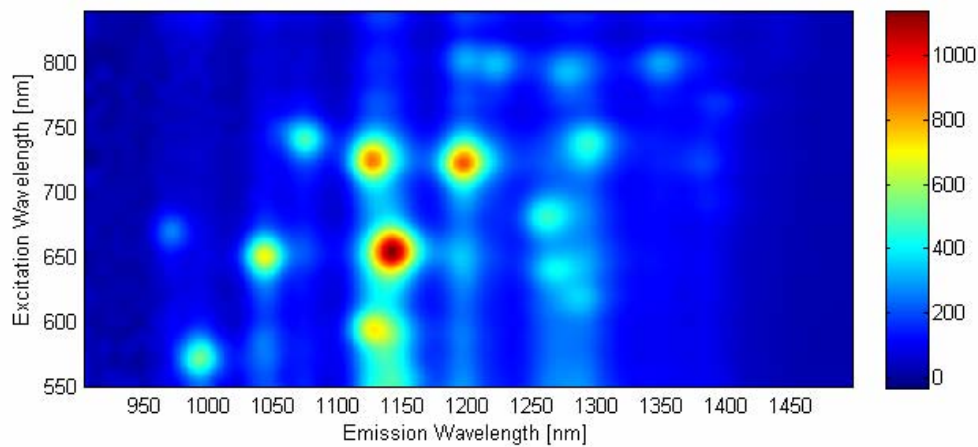

**Figure S10 | Fluorescence Excitation/Emission Spectrum.** Fluorescence excitation versus emission spectrum of SWNTs suspended in DSPE-mPEG. Although the sample shows SWNT excitation below 800 nm, an 808 nm laser was chosen for experiments to maximize the light penetration depth in tissues.

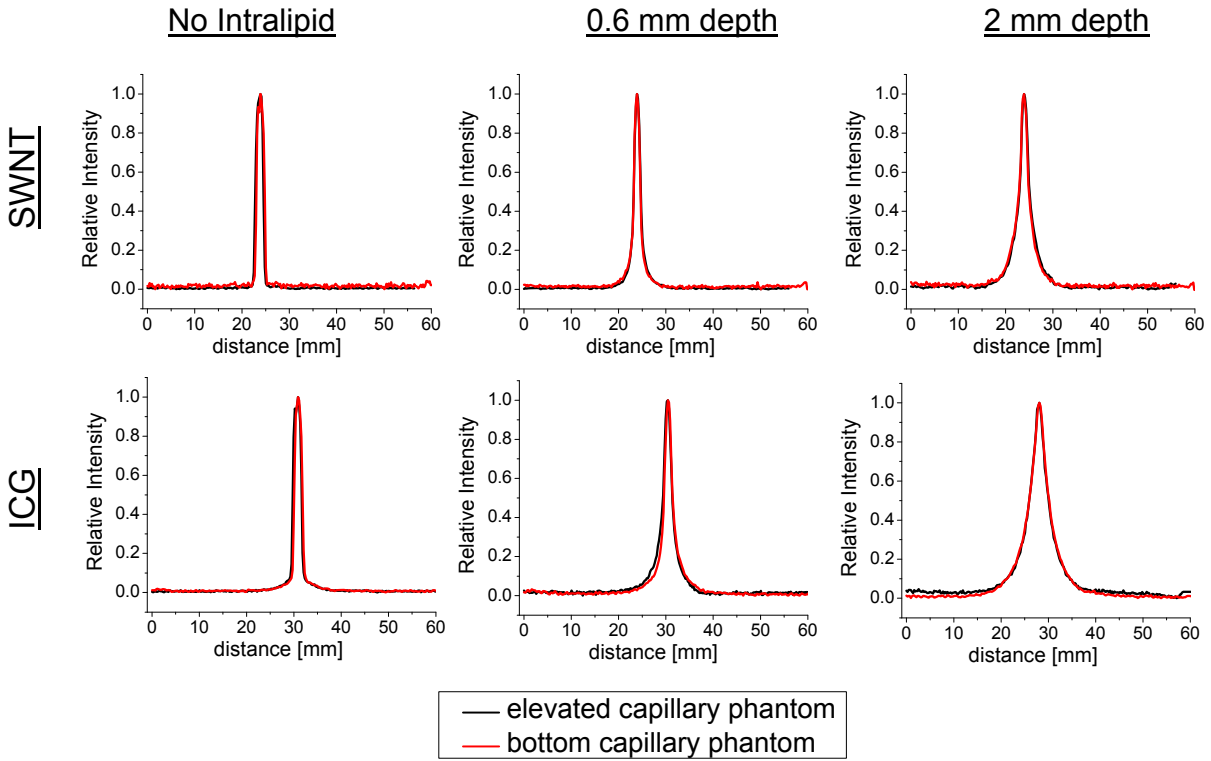

**Figure S11 | Capillary Phantom Set-Up.** To rule out signal alterations from the bottom of the dish during phantom tests, measurements were done in two geometries. The first geometry placed the capillary tube on the bottom of a glass dish, while the second elevated the capillary ~1.8 cm above the bottom of the dish. Both phantom geometries were imaged at multiple Intralipid® depths as described in the Methods section of the main text. Line cuts of both ICG and SWNT capillary images show that there is no difference in the signal scattering due to phantom geometry.

Video rate near infrared (NIR) video of back and side of mice 135 seconds following tail vein injection of SWNTs (Movie S1, S2 and S3) can be found online through PNAS.

**Movie S1 | Video of Intravenous Injection- Back View.** View of the back of a mouse during tail vein injection of SWNTs. The head of the animal is located at the top of screen. The video frame rate is 14 frames-per-second.

**Movie S2 | Video of Intravenous Injection- Side View.** View of the left side of a mouse during tail vein injection of SWNTs. The head of the animal is located at the top of screen. The video frame rate is 14 frames-per-second.

**Movie S3 | Video of Intravenous Injection- Breathing Removed.** View of the back of a mouse during tail vein injection of SWNTs. To remove major breathing movements that could affect PCA, frames where the animal is breathing have been removed. The head of the animal is located at the top of screen. The video frame rate is 14 frames-per-second.

## References for Supporting Information

1. Welsher, K., et al. (2009) A route to brightly fluorescent carbon nanotubes for near-infrared imaging in mice. *Nat Nano* 4: 773.
2. Lay, D. C. (2003) *Linear Algebra and Its Applications* (Addison Wesley, Boston).
3. Bachilo, S. M., et al. (2002) Structure-assigned optical spectra of single-walled carbon nanotubes. *Science* 298: 2361-236.
4. Kam, N. W. S., O'Connell, M., Wisdom, J. A. & Dai, H. (2005) Carbon nanotubes as multifunctional biological transporters and near-infrared agents for selective cancer cell destruction. *Proc Natl Acad Sci USA* 102: 11600-11605.
5. Splinter, R. & Hooper, B. A. (2007) *An introduction to biomedical optics* (Taylor & Francis, New York).
6. Vanstaveren, H. J., et al. (1991) Light-Scattering In Intralipid-10-Percent In The Wavelength Range Of 400-1100 Nm. *Appl Opt* 30: 4507-4514.